\documentclass[prl,twocolumn,floatfix]{revtex4}
\usepackage{graphicx}

\def\k{{\bf{k}}}

\begin{document}
\title{%
Phase diagram of the frustrated Hubbard model
}
\author{R.\ Zitzler}
\author{N.\ Tong}
\author{Th.\ Pruschke}
\author{R.\ Bulla}
\affiliation{Center for electronic correlations and magnetism,
Theoretical Physics III, Institute of Physics, University of Augsburg, 
86135 Augsburg, Germany}

\begin{abstract}
The Mott-Hubbard metal-insulator transition in the paramagnetic phase of
the one-band Hubbard model has long been used to describe similar
features in real materials like V$_2$O$_3$. Here we show that this 
transition is hidden inside a rather robust antiferromagnetic
insulator even in the presence of comparatively strong magnetic frustration.
This result raises the question of the relevance of the Mott-Hubbard metal-insulator
transition for the generic phase diagram of the one-band Hubbard model.
\end{abstract}
\maketitle              

The microscopic description of magnetism and metal-insulator transitions  constitutes
one of the major research activities in modern solid state theory. Especially transition
metal compounds like V$_2$O$_3$, LaTiO$_3$, NiS$_{2-x}$Se$_{x}$ and the cuprates
show metal-insulator transitions and magnetic order depending on composition, pressure
or other control parameters \cite{imada98}. The paramagnetic insulating phase observed in
these materials is believed to be a so-called Mott-Hubbard insulator
due to electron-electron correlations; in contrast to Slater or band insulators like
SrTiO$_3$.

The simplest model showing both magnetism and a correlation-induced metal-insulator
transition (MIT) is the one-band Hubbard model \cite{hubbard}
\begin{equation}
\label{equ:hubbard}
H=-\sum_{i,j,\sigma}t_{ij}
c^{\dagger}_{i\sigma}c^{\phantom{\dagger}}_{j\sigma}
+\frac{U}{2}\sum_{i\sigma}n_{i\sigma}n_{i\bar{\sigma}}\;\;.
\end{equation}
Considerable progress in understanding the physics of this simple but nevertheless
non-trivial model has been achieved in the last decade through the development of
the dynamical mean-field theory (DMFT) \cite{mv,pradv,rmp}.
In particular, the phase diagram for
the unfrustrated Hubbard model is very well understood \cite{pradv,rmp}. 
At half-filling the physics is dominated by an antiferromagnetic insulating
phase (AFI) for all $U>0$ with a maximum $T_N\approx0.15W$ around $U\approx W$, where $W$ is
the bandwidth of the non-interacting system. 
For finite doping, the antiferromagnetic phase persists up to a critical
doping $\delta_c$ \cite{jazph} and in addition shows phase separation \cite{pvd,zitz1}.
For very large values of $U$ the antiferromagnetic phase is replaced
by a small region of Nagaoka type ferromagnetism \cite{nagoka,oberm,ulmke}.

Under rather special assumptions \cite{muha89}, it is possible to introduce complete
magnetic frustration and suppress the antiferromagnetic phase. In this case, a
transition from a paramagnetic metal (PM) to a paramagnetic insulator
(PI) becomes
visible at half filling. At $T=0$ it occurs at a  value of the
Coulomb parameter $U_c\approx 1.5W$ \cite{jazph,rmp,bulprl}. Interestingly, the
transition is of first order \cite{rmp,bulcosvol} for $T>0$ with a second order
end point at a $T_c\approx 0.017W$ and $U_c\approx 1.2W$. Note that $T_c\ll T_N^{\rm max}$.

A closer look at the phase diagram of  V$_2$O$_3$ \cite{mcwhan} reveals a 
strikingly similar scenario,
and indeed the DMFT results for the Hubbard model have been used as
a qualitative explanation \cite{rmp,roz95}.
For a proper description of this  material, however, the antiferromagnetic
phase below $T_N\approx160$K \cite{mcwhan} has to be taken into account.
It was argued and generally accepted \cite{rmp} that the introduction of
partial magnetic frustration will lead to the anticipated situation, where the
MIT extends beyond the antiferromagnetic phase at low temperatures.
The merging of these two transitions presents an interesting problem on its own,
because it is commonly believed that the magnetic transition should
be of second order. Furthermore, previous results for a system with magnetic
frustration show, in addition to the desired effect of
reducing $T_N$, an extended antiferromagnetic metallic (AFM) phase at
small $U$, preceeding the transition to the AFI \cite{rmp,chitra}.
This observation suggests an appealing possibility to link the MIT in the paramagnetic phase
with a transition from an AFM to an AFI.

In this paper, we present results from a calculation using Wilson's numerical
renormalization group approach (NRG) \cite{nrg} and exact diagonalization
techniques (ED) \cite{krauth} to solve the DMFT equations \cite{rmp,bulprhew} for the
Hubbard model with magnetic frustration at half filling. We show that frustration of
the magnetic interactions through incorporation of suitable long-range hopping
 does {\em not} lead to the previously reported sequence
PM$\leftrightarrow$AFM$\leftrightarrow$AFI with an extended region of
an AFM at $T=0$
\cite{rmp,chitra}. Instead, we observe a first-order 
 transition PM$\leftrightarrow$AFI.
Furthermore, the reduction of $T_N$ is too small to result in the qualitative
phase diagram of V$_2$O$_3$.

The natural choice for studying the effect of magnetic frustration is the simple
hypercubic lattice with nearest and next-nearest neighbor hopping. In the DMFT, the lattice
structure only enters via the dispersion of the noninteracting band states, and the
corresponding $\k$ sums can conveniently be transformed into energy integrals
using the free density of states (DOS) \cite{rmp}. A further simplification arises
if one considers lattices in the limit of large coordination number. Especially for the
simple hypercubic lattice the DOS then becomes a Gaussian \cite{mv,muha89} and the integrals
can be performed analytically \cite{pradv}. 

The investigation of magnetic properties is
straightforward, too. In the case of the N\'eel state,
the lattice is divided into A and B sublattices which results in a matrix
structure of the DMFT equations \cite{rmp}. 
An antiferromagnetic N\'eel order then corresponds to
a finite staggered magnetization $m_S>0$ with $m_A=m_S$ and $m_B=-m_S$. 
Unfortunately, the Gaussian DOS of the hypercubic lattice has no real
band edges, but stretches to infinity; the resulting exponential tails 
therefore prevent
a clear distinction between metal and insulator at $T=0$, as has been observed in a Hartree calculation
for the hypercubic lattice with next-nearest neighbor hopping \cite{hofvol}. 

Georges et al.\ \cite{rmp} suggested an extension of the DMFT equations
for the Bethe lattice which introduces magnetic frustration with the DOS
having finite support; this results in an analytically  tractable form of the 
DMFT equations even for the AB-lattice.

For the standard Bethe lattice with nearest-neighbor hopping $t$ and in the limit of infinite coordination
number the DMFT equations on an AB-lattice acquire the form \cite{rmp}
\begin{equation}\label{equ:bethedmft}
\begin{array}{l}\displaystyle
G_{A\sigma}(z)=\frac{1}{z+\mu-\Sigma_{A\sigma}(z)-\frac{t^2}{4}G_{B\sigma}(z)}
 \ , \\[5mm]
\displaystyle
G_{B\sigma}(z)=\frac{1}{z+\mu-\Sigma_{B\sigma}(z)-\frac{t^2}{4}G_{A\sigma}(z)} \ .
\end{array}
\end{equation}
Frustration is then introduced via additional terms in the denominators of (\ref{equ:bethedmft}) \cite{rmp,chitra}
\begin{equation}\label{equ:bethedmft2}
\begin{array}{l}\displaystyle
G_{A\sigma}(z)=\frac{1}{z+\mu-\Sigma_{A\sigma}(z)-\frac{t_1^2}{4}G_{B\sigma}(z)-\frac{t_2^2}{4}G_{A\sigma}(z)} \ , \\[5mm]
\displaystyle
G_{B\sigma}(z)=\frac{1}{z+\mu-\Sigma_{B\sigma}(z)-\frac{t_1^2}{4}G_{A\sigma}(z)-\frac{t_2^2}{4}G_{B\sigma}(z)} \ ,
\end{array}
\end{equation}
and the constraint $t^2=t_1^2+t_2^2$. In the paramagnetic case, the equations
(\ref{equ:bethedmft2}) reduce to those of a standard Bethe lattice which,
for example, result in the well-studied Mott-transition. 
Furthermore, despite the frustration introduced, the system
is still particle-hole symmetric. Especially for half filling this feature reduces the numerical effort
quite drastically. Note that a similar suggestion exists for the hypercubic lattice \cite{muha89}, too.

Invoking the symmetry $G_{A\sigma}(z)=G_{B\bar\sigma}(z)$ valid for the N\'eel state, eqs.\ (\ref{equ:bethedmft2})
reduce to two coupled nonlinear equations which we solve iteratively. In the course of the iterations,
the quantity $\Sigma_{\sigma}(z)$ has to be calculated from the solution of a generalized single impurity Anderson
model \cite{rmp}. For that task we employ the NRG \cite{nrg}, suitably extended to treat spin-polarized
situations \cite{zitz1,hof01}. 

\begin{figure}[htb]
\begin{center}
\includegraphics[width=0.435\textwidth,clip]{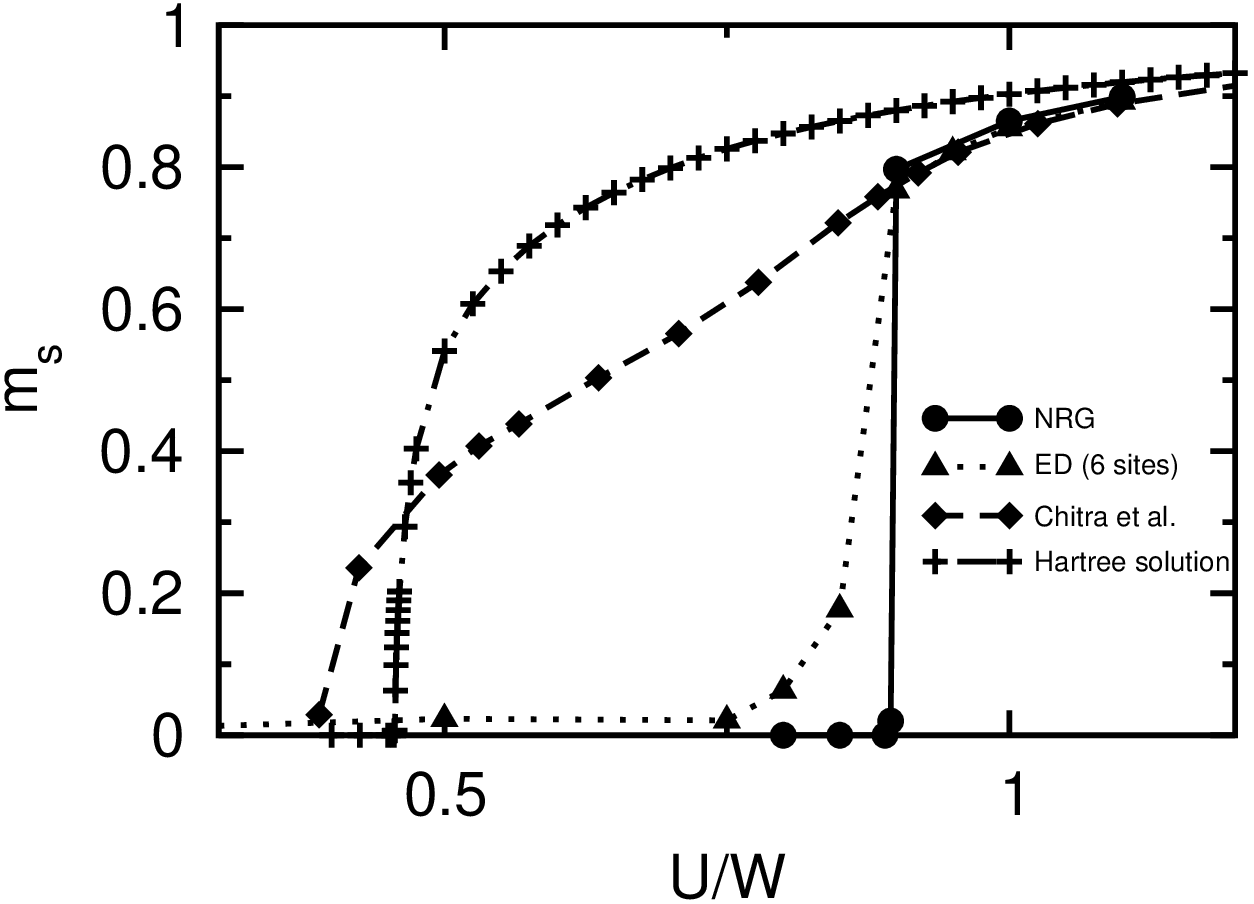}
\end{center}
\caption[]{Staggered magnetization $m_S$ as function of $U$ at $T=0$. 
The circles are the results from
NRG; the triangles from an ED calculation for 6 sites, while the diamonds are taken from
ref.~\onlinecite{chitra}. For comparison, the results of a Hartree calculation are given by the
crosses.\label{fig:groundstate}}
\end{figure}

Let us first discuss the results for the magnetization as function of $U$ for $T=0$.
In the following, we fix $t_2/t_1=1/\sqrt{3}\approx 0.58$ \cite{comment} and use the bandwidth $W$ of the non-interacting
system as our energy scale. The NRG results in Fig.~\ref{fig:groundstate} (circles)
show a completely different behavior as compared to
 the ED data (diamonds) from ref.~\onlinecite{chitra}.
Instead of a continuous increase of the staggered magnetization $m_S$ for $U>U_c\approx0.4W$ as suggested by both a
Hartree calculation
(crosses) and the data from ref.~\onlinecite{chitra}, we find a jump in $m_S$ at a considerably larger $U_c\approx0.9W$.
To clarify
this discrepancy, we performed ED calculations, resulting in the triangles shown in Fig.~\ref{fig:groundstate}.
We find good agreement with our NRG results, the transition systematically approaching the NRG curve
with increasing size of the system diagonalized in the ED procedure. We furthermore observed a rather strong
dependence of the ED results on details of the numerical procedure, especially on the energy cutoff introduced
in calculating $G(z)$.
Decreasing this cutoff systematically shifts the ED result towards the one found in \cite{chitra}. The NRG,
on the other hand, is stable with respect to changes in the parameters controlling its
numerical accuracy.
\begin{figure}[htb]
\begin{center}
\includegraphics[width=0.435\textwidth,clip]{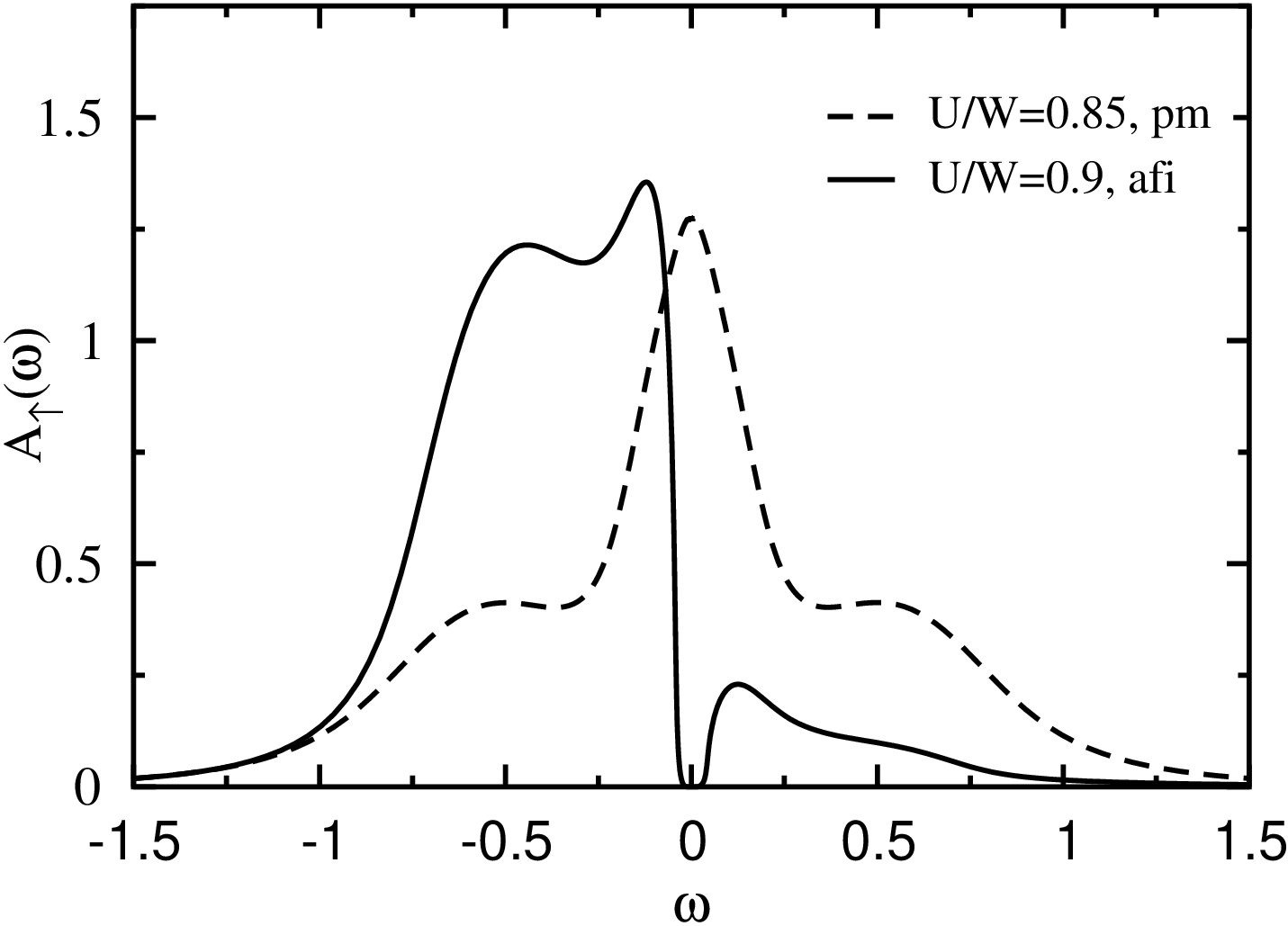}
\end{center}
\caption[]{Density of states
 for spin up on an A lattice site as function of frequency.\label{fig:DOS1}}
\end{figure}

Another important question is the existence of an antiferromagnetic {\em metallic} solution 
of the DMFT equations.
Figure~\ref{fig:DOS1} shows the NRG results for the DOS for $T=0$ and
spin up on an A lattice site. Due to particle-hole symmetry
the DOS for spin down on A sites (or spin up on B sites) can be obtained by $\omega\to-\omega$.
The full and dashed lines represent the AFI solution 
for $U\searrow U_c$ and the PM solution for
$U\nearrow U_c$, respectively. Clearly, the magnetic solution is insulating with a well-developed gap at the Fermi
energy. Quite generally, we were not able to find a stable AFM solution at $T=0$.

The discontinuity in the staggered magnetization $m_S$ at the transition PM$\leftrightarrow$AFI implies 
a first order transition and the existence of a hysteresis region.
\begin{figure}[htb]
\begin{center}
\includegraphics[width=0.435\textwidth,clip]{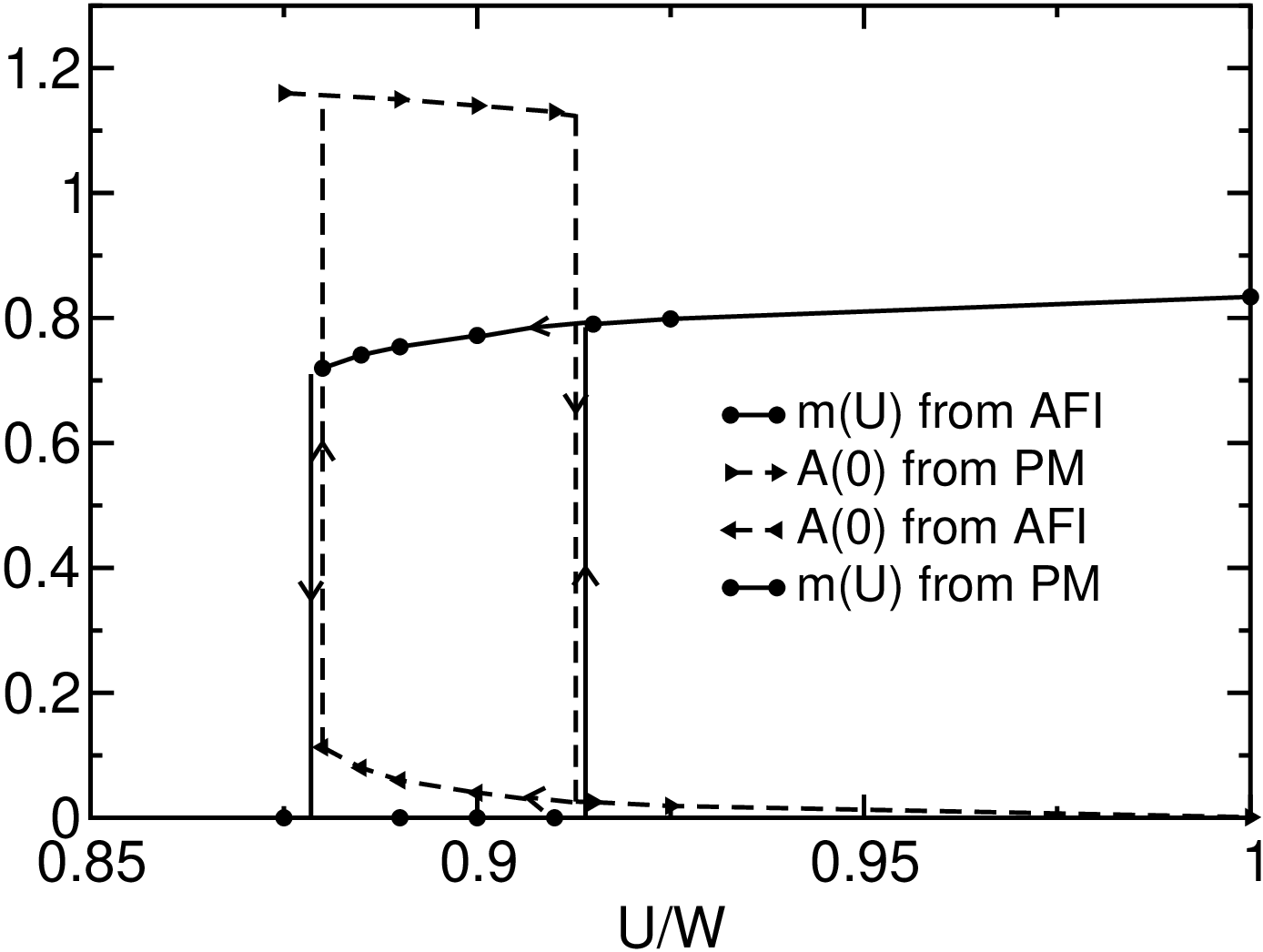}
\end{center}
\caption[]{Staggered magnetization (solid lines) and total DOS at the Fermi energy 
(dashed lines)
as function of $U$ in the vicinity of $U_c$ for $T=0.0155W$. 
The arrows indicate that the DMFT solutions have been obtained by either increasing
$U$ ($\rightarrow$) or decreasing
$U$ ($\leftarrow$).
\label{fig:hysteresis}}
\end{figure}
Indeed, starting from the paramagnet at $U\ll U_c$ and increasing $U$ results in a magnetization curve
different from the one obtained by starting at $U\gg U_c$ and decreasing $U$. This is apparent from
Fig.~\ref{fig:hysteresis}
(full lines) where a region of hysteresis can be observed in the staggered magnetization
(for temperature
$T=0.0155W$). At the same time the total DOS at the Fermi energy $A(0)=A_\uparrow(0)+A_\downarrow(0)$ shows
hysteresis between metallic and insulating behavior in exactly the same $U$ region. Note, that due to the
finite temperature the DOS at the Fermi level is not exactly zero in the N\'eel
state, but strongly reduced as compared to the metal
\cite{bulcosvol}. 

It is of course important to verify that the hysteresis found for small $U$ is not 
some kind of artefact.
This can most conveniently be shown by looking at the transition at large $U$. Due
\begin{figure}[htb]
  \begin{center}
    \includegraphics[width=0.435\textwidth,clip]{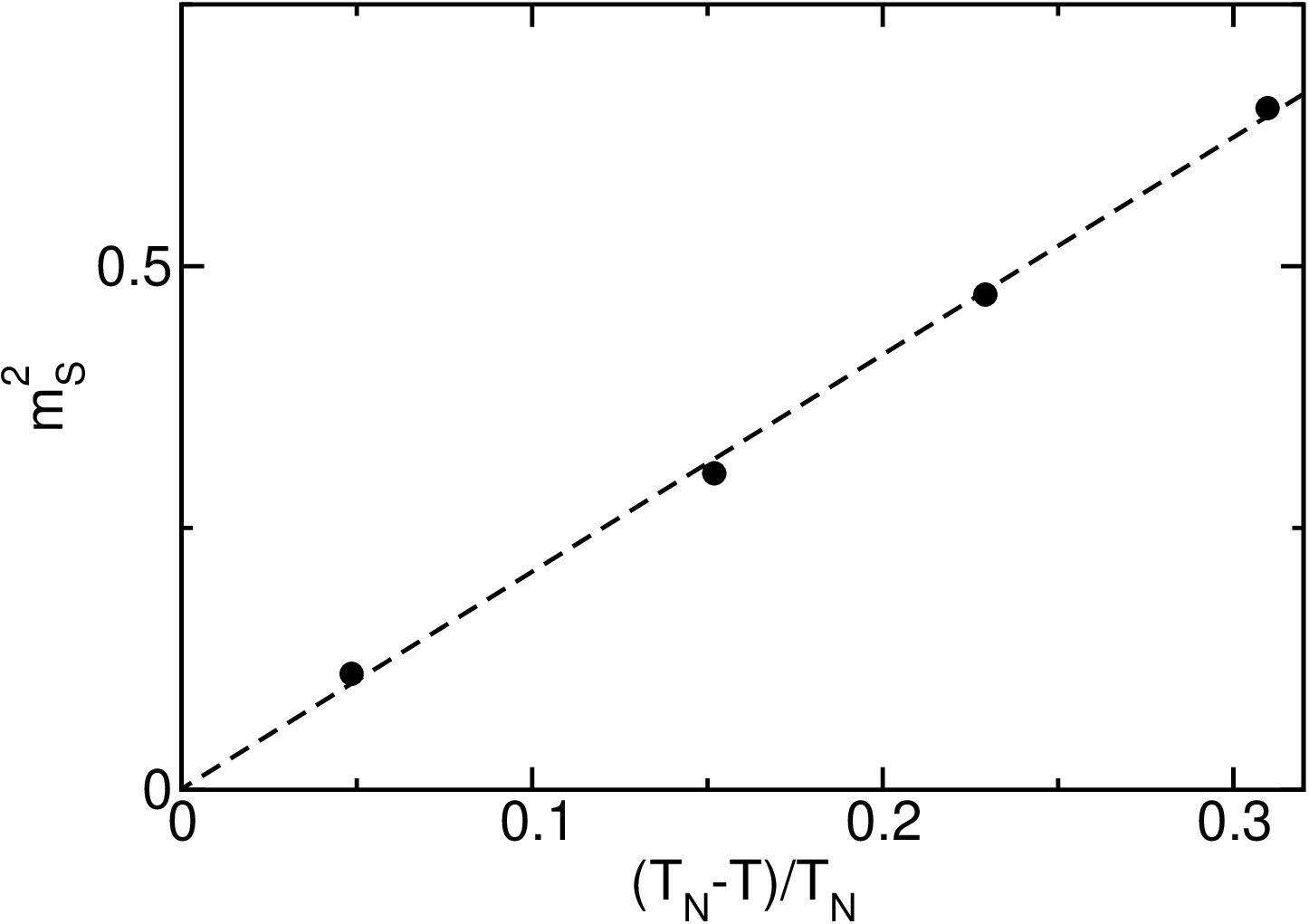}
  \end{center}
\caption[]{Squared staggered magnetization $m_S$ as function of $T$ 
at $U/W=2$. Note that $m_S^2$ vanishes continously like $T_N-T$ as
$T\searrow T_N$.\label{fig:largeU}}
\end{figure}
to the mapping of the Hubbard model to a Heisenberg model in this regime, one should expect the transition
to be of second order, with the staggered magnetization vanishing continously
like $m_S\propto\sqrt{T_N-T}$ when approaching $T_N$ from below.
That this is indeed the case is apparent from Fig.~\ref{fig:largeU}, where
we show the squared staggered magnetization as function of $T$ for
$U/W=2$. The transition is thus of second order with the expected mean-field
exponent in this region of the phase diagram.

\begin{figure}[htb]
  \begin{center}
    \includegraphics[width=0.435\textwidth,clip]{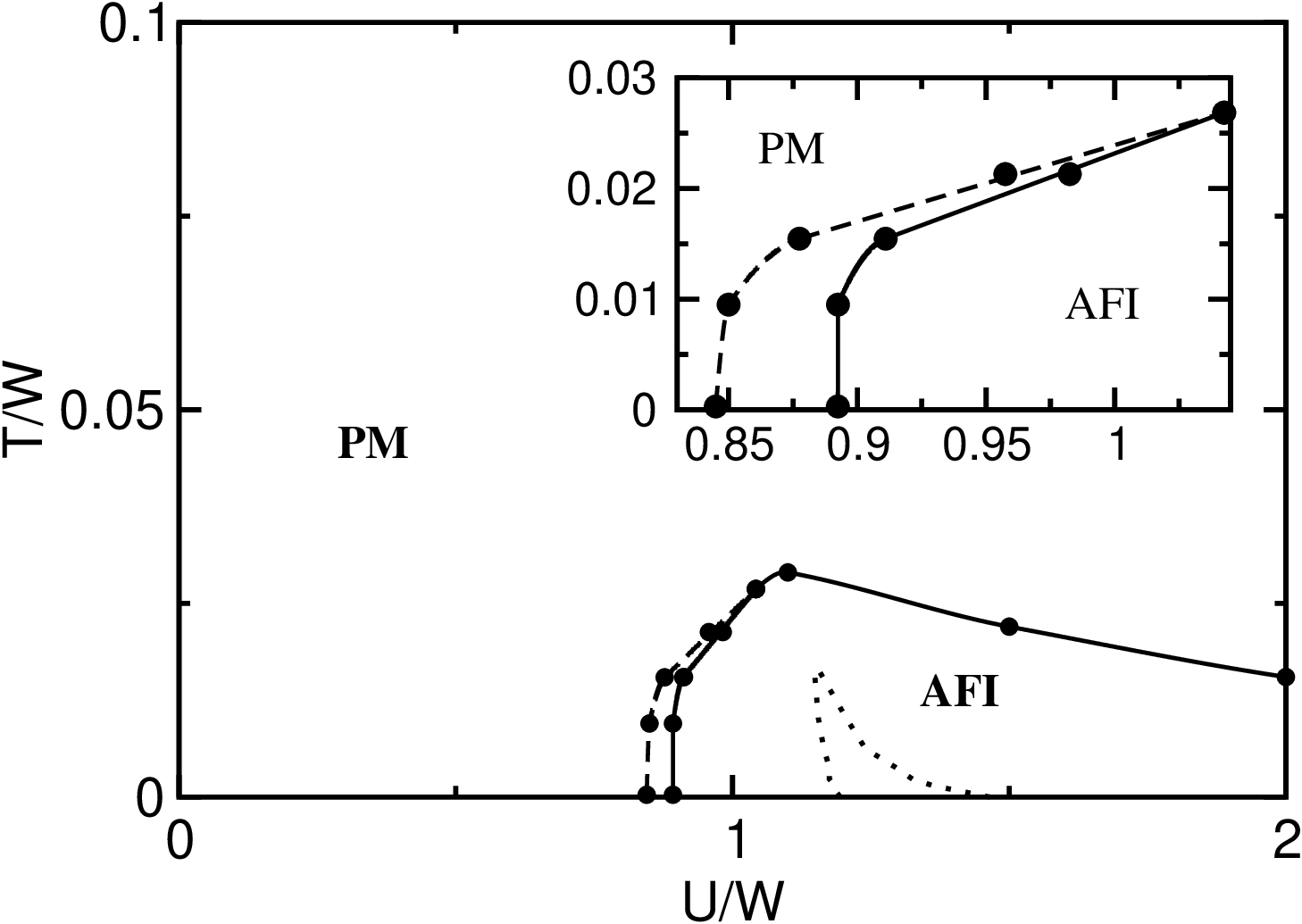}
  \end{center}
\caption[]{Magnetic phase diagram for the Hubbard model with frustration
as defined by eqs.~(\ref{equ:bethedmft2}) and $t_2/t_1=1/\sqrt{3}$. The dotted lines inside the AFI
denote the coexistence region for the paramagnetic MIT.
The inset shows an enlarged view of the region with coexistence of 
PM and AFI.\label{fig:phasediagram}}
\end{figure}

Collecting the results for the transitions and the hysteresis region for different
temperatures leads to the phase diagram in Fig.~\ref{fig:phasediagram}.
An enlarged view of the region showing coexistence of PM and AFI
is given in the inset, where the full line represents the
transition PM$\rightarrow$AFI with increasing $U$ and the dashed line
the transition AFI$\rightarrow$PM with
decreasing $U$. These two lines seem to merge at a value of $U\approx W$
for this particular value of $t_2$, with a critical temperature for this
endpoint $T_c\approx0.02W$.
Note that, even in the presence of such a sizeable $t_2$, the antiferromagnetic
phase
still completely encompasses the paramagnetic MIT (dotted lines in the main panel
of Fig.~\ref{fig:phasediagram} \cite{bulcosvol}). 

It is, of course, interesting to see how the magnetic phase evolves with
increasing $t_2$ and in particular how its boundary crosses the
paramagnetic MIT. We find that increasing $t_2$ does not change the form of the magnetic
\begin{figure}[htb]
  \begin{center}
    \includegraphics[width=0.435\textwidth,clip]{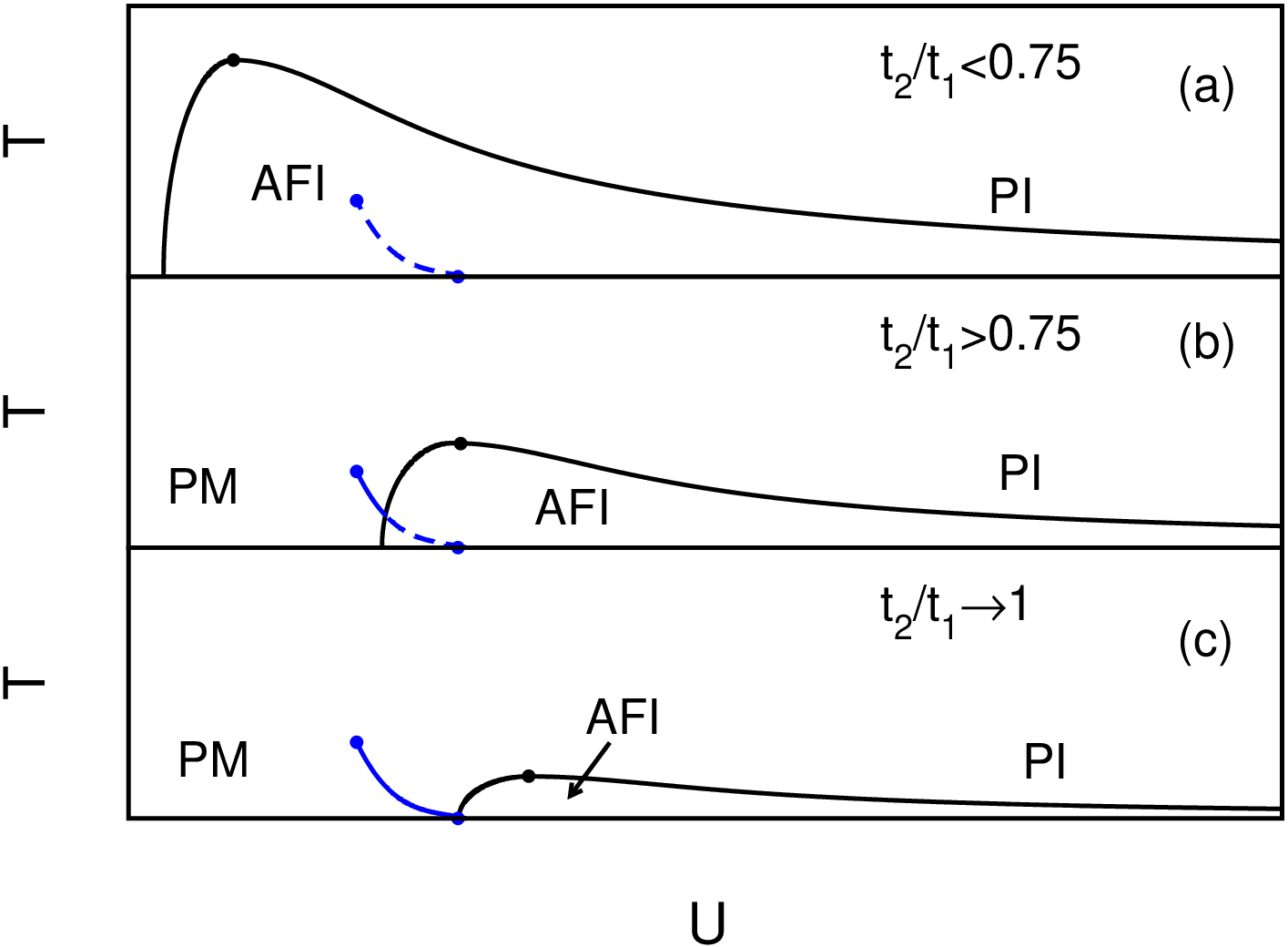}
  \end{center}
\caption[]{Schematic evolution of the magnetic phase diagram with
increasing frustration. The dots on the phase transition lines
denote the critical endpoints of the first order transitions.\label{fig:pd_predicted}}
\end{figure}
phase in Fig.~\ref{fig:phasediagram} qualitatively, but mainly shifts
the lower critical $U$ and decreases the maximum $T_N$. The calculated
estimates for those two quantities as function of $t_2$ lead to the
schematic evolution of the phase diagram presented in Fig.~\ref{fig:pd_predicted}a-c.
Here, only the true phase boundaries are shown.
A direct calculation of the free energy at finite temperatures is presently
not possible with the NRG method, so we cannot calculate the actual transition
line separating the paramagnetic and AF phases. The transition lines in Fig.~\ref{fig:pd_predicted}a-c are
therefore a guide to the eye only. For the Mott transition, the
position of the actual transition line has been calculated in ref.~\onlinecite{tong}.

Figure~\ref{fig:pd_predicted}a shows the qualitative phase diagram
corresponding to Fig.~\ref{fig:phasediagram} with the line
of first order transitions ending in a critical point. 
Upon further increasing the value of $t_2$, the first order transition
lines from both the PM$\leftrightarrow$AFI and the Mott transition cross
(Fig.~\ref{fig:pd_predicted}b), thus exposing a finite region of the Mott
insulator and a transition PI$\leftrightarrow$AFI.
Finally, for even higher values of $t_2$, the PM$\leftrightarrow$AFI
transition at $T=0$ approaches the Mott transition and
$T_N$ is reduced significantly (Fig.~\ref{fig:pd_predicted}c). Note that in the limiting case
$t_2 = t_1$ the AFI phase completely vanishes due to the structure of
the DMFT equations (\ref{equ:bethedmft2}). However, for $t_2
\rightarrow t_1$ there is always a finite
antiferromagnetic exchange $J\propto (t_1^2-t_2^2)/U$ which is
sufficient to stabilize an antiferromagnetic
ground state for $U>U_c$ of the Mott transition.

From these results we conclude that frustration as introduced via 
eqs.~(\ref{equ:bethedmft2})
is not sufficient to qualitatively reproduce the
phase diagram of materials like\ V$_2$O$_3$. In particular, the Mott transition extends
beyond the AFI region only for unphysically large values of $t_2$.

The question remains whether it is possible at all to reproduce qualitatively
the scenario observed in V$_2$O$_3$ within a one-band model. 
Based on our results reported here, we rather believe that one has to take
into account additional degrees of freedom, for example phonons (within a Holstein-Hubbard model)
or orbital degeneracies (within a multi-band Hubbard model).

\begin{acknowledgments}
We acknowledge useful conversations with
M.\ Vojta,
G.\ Kotliar,
R.\ Chitra
and
P.J.G.\ van Dongen.
This work was supported by the DFG through the collaborative research center SFB 484, the
Leibniz Computer center and the Computer center of the Max-Planck-Gesellschaft in Garching.
NT acknowledges the support by the Alexander von Humboldt foundation.

\end{acknowledgments}


\begin{thebibliography}{99}
\bibitem{comment} This value is the same as the one used in \cite{rmp,chitra}
\end{thebibliography}
\end{document}